\documentclass[aps,prd,twocolumn,superscriptaddress,nofootinbib]{revtex4-2}

\usepackage{amsmath,amssymb,bm}
\usepackage{graphicx}
\usepackage{hyperref}
\usepackage{url}
\usepackage{xcolor}

\newcommand{\red}[1]{\textcolor{black}{#1}}

\begin{document}

\title{Universal Mechanism of Dynamical Memory in Light-Nuclei Formation and Cosmological Bose–Einstein Condensation}

\author{Takeshi Fukuyama}
\affiliation{Research Center for Nuclear Physics (RCNP), Osaka University, Ibaraki, Osaka 567-0047, Japan}

\date{\today}

\begin{abstract}
We identify a common reduced nonequilibrium structure in
light-nuclei formation in heavy-ion collisions and
cosmological Bose--Einstein condensation.  In the
heavy-ion system, eliminating a finite-lived intermediate
hadronic sector, represented minimally by the $\Delta$
component, generates an exponential memory kernel.  In
the Fukuyama--Morikawa--Tatekawa (FMT) cosmological
system, the localized collapse component is generated
from the coherent field and feeds back through the
expansion rate, producing the causal sequence
$\rho_\phi\to\rho_l\to H\to\rho_\phi$.  Eliminating this
additional dynamical sector likewise generates a
history-dependent reduced equation.  Although the two
memory kernels and their microscopic origins are
different, they realize the same generic causal structure:
elimination of a dynamically relevant coupled sector
leaves a memory of its previous evolution.  This provides
a precise sense in which the two systems belong to the
same reduced-dynamical universality class and leads to a
time-scale criterion separating approximately Markovian
from genuinely non-Markovian evolution.

As a concrete application, we consider hypertriton
formation through direct $p+n+\Lambda$ and sequential
$d+\Lambda$ channels.  In the minimal reduced
description, the direct channel acts as a local source,
whereas elimination of the intermediate deuteron sector
generates a retarded sequential contribution.  The
resulting prediction is that the sequential-to-direct
ratio $R_{\rm seq/dir}$ should depend on the duration and
expansion history of the hadronic stage, with
$R_{\rm seq/dir}$ increasing with the effective evolution
time when the sequential contribution retains its
correlations more efficiently.  Existing coalescence
calculations provide a quantitative benchmark in which
this ratio changes by about $40\%$ while the total
hypertriton abundance changes by only about $20\%$.
More generally, several independent relaxation modes in
the eliminated sector generate a multi-scale memory
kernel.  These results show that equilibrium-like final
abundances can conceal dynamical information that remains
encoded in causal memory and in the relative weights of
competing formation pathways.
\end{abstract}

\maketitle

\section{Introduction}

The formation of fragile bound or coherent structures in rapidly
evolving nonequilibrium environments is a general problem that
appears in widely separated areas of physics. A final low-energy
structure may emerge under conditions very different from those
prevailing during the earlier stages of the evolution.
Consequently, an apparently thermal or equilibrium-like final state
does not by itself determine the dynamical path by which that state
was reached. Intermediate or auxiliary dynamical sectors may store
correlations and information about the previous evolution even when
they are no longer explicitly visible in the final observables.

Light-nuclei production in relativistic heavy-ion collisions provides
an important example. Loosely bound nuclei such as the deuteron,
with binding energy $E_{\rm bind}\simeq 2.2~\mathrm{MeV}$, are
observed in collisions whose characteristic temperatures are much
larger than their binding energies. Nevertheless,
statistical-hadronization descriptions reproduce many of the measured
light-nuclei yields remarkably well. This long-standing ``thermal
paradox'' has motivated several interpretations, including
statistical hadronization, late-stage coalescence, and dynamical
descriptions of the hadronic evolution
\cite{ALICEnews2025,ALICE502,STAR2022,ALICE2025Nature,
BegunFlorkowski2015}.
Recent ALICE measurements and correlation studies provide further
motivation for examining the role of hadronic resonances and
finite-lifetime correlation channels during the late evolution of
the system.

In the minimal heavy-ion description considered here, a finite-lived
$\Delta$ component is used as a representative intermediate hadronic
sector. Its purpose is not to identify a unique microscopic origin
of deuteron production. Rather, it provides a simple realization of
a more general dynamical question: what remains in the effective
evolution of the retained variables when a finite-response-time
intermediate sector is eliminated? As shown below, eliminating the
$\Delta$ component produces an explicitly retarded contribution to
the reduced nucleon dynamics.

A seemingly very different realization of the same question occurs
in cosmological Bose--Einstein condensation (BEC) of scalar-field
dark matter. In the Fukuyama--Morikawa--Tatekawa (FMT) scenario
\cite{FMT2008,FM2009,Fuku2025}, the coarse-grained dynamics contains
a boson-gas component $\rho_g$, a coherent component $\rho_\phi$,
and a localized collapse-induced component $\rho_l$. Their dominant
transfer topology is
\[
 \rho_g\longrightarrow\rho_\phi,
 \qquad
 \rho_\phi\longrightarrow\rho_l .
\]
Thus the localized component is not a microscopic precursor to
condensation and should not be identified with the intermediate
$\Delta$ sector of the heavy-ion example. Nevertheless, $\rho_l$
contributes to the Friedmann equation, while the resulting Hubble
rate enters the evolution of $\rho_\phi$. The full dynamics therefore
contains the feedback loop
\[
 \rho_\phi\longrightarrow\rho_l
 \longrightarrow H\longrightarrow\rho_\phi .
\]
Eliminating the localized sector consequently leaves a history
dependence in the reduced coherent dynamics.

These two systems have very different microscopic degrees of freedom,
interactions, characteristic time scales, and expansion dynamics.
The universality considered in this work does not refer to these
microscopic ingredients, nor does it require an identical cascade
topology. It concerns instead the structure of the reduced dynamics
obtained after a dynamically relevant additional sector has been
eliminated. In both examples the resulting evolution can be written
in the generic causal form
\[
 \dot Y(t)
 =
 F[Y(t)]
 +
 \int_{t_0}^{t}dt'\,
 {\cal K}(t,t')\,{\cal S}[Y(t')],
\]
where the memory kernel ${\cal K}$ is determined by the particular
physical realization. The same general structure arises naturally
in the real-time Schwinger--Keldysh reduction developed in
Ref.~\cite{Fuku2026}.

The purpose of the present work is to demonstrate explicitly how this
common reduced-dynamical structure emerges in the two physically
distinct systems. In the heavy-ion example, elimination of the
finite-lived $\Delta$ component produces an exponential memory kernel
controlled by $\Gamma_\Delta^{-1}$. In the FMT system, elimination
of the localized component, together with its gravitational feedback
through the Friedmann equation, produces a different memory kernel
in the coherent dynamics. The kernels need not have the same
functional form. Their difference encodes the microscopic physics;
the causal reduced structure that survives the elimination is what
is universal.

The comparison also provides a simple time-scale criterion. If the
response time of the eliminated sector is much shorter than the
characteristic evolution time of the retained variables, its effect
becomes approximately local and a Markovian description is adequate.
If the response time is not negligible, the previous evolution of
the eliminated sector survives as memory in the effective dynamics.
This criterion is independent of whether the underlying system is
hadronic or cosmological.
The reduced-dynamical viewpoint also leads to predictions
beyond these two realizations. If several dynamically
relevant intermediate sectors are eliminated, the resulting
memory kernel generally contains several characteristic
relaxation scales, reflecting the relaxation spectrum of
the eliminated dynamics. The existence of several reaction
pathways alone, however, does not imply several independent
relaxation modes. As a concrete application, we consider
hypertriton ${}^{3}_{\Lambda}{\rm H}$ formation, for which
direct $p+n+\Lambda$ coalescence and a sequential channel
involving an intermediate deuteron provide distinct
formation pathways. In the minimal reduced description,
the direct channel acts as a local source, whereas
eliminating the intermediate deuteron sector generates a
retarded contribution to the sequential channel. The
resulting prediction is that the relative weight
$R_{\rm seq/dir}$ of sequential to direct formation should
depend on the duration and expansion history of the
hadronic stage. In the regime in which the sequential
contribution retains its correlations more efficiently
than the direct contribution, one expects
\[
 \tau_{\rm macro}\uparrow
 \quad\Longrightarrow\quad
 R_{\rm seq/dir}\uparrow .
\]
The calculation of Ref.~\cite{ZhangKo2018} provides a
quantitative benchmark for the possible size of this
effect: over an evolution interval of about
$35~{\rm fm}/c$, the total hypertriton abundance changes
by only about $20\%$, whereas the sequential-to-direct
ratio changes by about $40\%$. This numerical value is
used here as a benchmark rather than as an absolute
prediction of the present reduced model.

The remainder of this paper is organized as follows.
Section~II introduces the minimal reservoir-mediated
description of light-nuclei formation and illustrates its
time ordering numerically. Section~III reviews the reduced
cosmological BEC description and the original three-component
FMT dynamics. Section~IV contains the central theoretical
development of this work. We first derive the generic
causal reduced equation obtained by eliminating an
additional dynamical sector and show explicitly how it is
realized in the heavy-ion and FMT systems. In the
heavy-ion case the memory kernel is generated directly by
the finite lifetime of the intermediate hadronic sector,
whereas in the FMT case it arises indirectly through the
gravitational feedback
$\rho_\phi\to\rho_l\to H\to\rho_\phi$.
We then formulate the time-scale criterion separating the
approximately Markovian regime from the regime in which
memory must be retained. The reduction is subsequently
generalized to several intermediate dynamical sectors,
showing that their relaxation spectrum is encoded in a
multi-scale memory kernel. Finally, hypertriton formation
is considered as a concrete application: elimination of
the intermediate deuteron sector produces a retarded
sequential contribution and leads to a testable
evolution-time dependence of the relative formation-channel
composition. Section~V summarizes the discussion and
conclusions.

\section{Reservoir-mediated light-nuclei formation
in heavy-ion collisions}

\subsection{Time-ordered nonequilibrium cascade}

We first consider the heavy-ion realization of the general
reservoir-mediated structure introduced above.  By a
``time-ordered nonequilibrium cascade'' we mean a dynamical process
in which correlations relevant for a fragile final state are generated
and redistributed through intermediate degrees of freedom before the
final bound state can survive.

The relevant microscopic hadronic dynamics may involve nucleons,
pions, baryonic resonances, and continuum correlation channels.
Rather than identifying a unique microscopic precursor of the
deuteron, we collectively regard finite-lifetime intermediate
correlations as a reservoir $R$.  The essential point is then a
competition of time scales.  At early times, when the hadronic medium
is hot and dense, formation and destruction processes prevent a
fragile bound state from surviving efficiently.  During the subsequent
expansion and cooling, intermediate hadronic correlations evolve and
can feed nucleons into regions of phase space in which deuteron
breakup is less effective.

The term ``time ordered'' refers to this sequence,
\[
 {\rm hadronic\ correlations}
 \;\longrightarrow\;
 {\rm reservoir}
 \;\longrightarrow\;
 {\rm light\ nucleus}.
\]
The intermediate reservoir need not correspond to a single particle
species.  In the minimal realization considered below, the $\Delta$
resonance is used as a representative finite-lifetime component,
while more general pion--nucleon and pion--deuteron correlation
channels may contribute in a realistic hadronic system.  Thus the
role assigned to the $\Delta$ below is illustrative of the reservoir
mechanism and is not a claim that $\Delta$ decay constitutes the
unique microscopic origin of deuteron production.

\subsection{From the thermal paradox to a reservoir picture}

Statistical models treat light nuclei as additional species in a
grand-canonical ensemble at chemical freeze-out.  The empirical
success of such descriptions is striking because nuclear binding
energies are much smaller than the characteristic temperatures of
the hadronic system.  This observation has motivated different
interpretations of when and how the finally observed nuclei are
formed \cite{GG2021}.

In a coalescence description, light nuclei are formed from nucleons
that are sufficiently close in phase space at a late stage of the
evolution.  In its simplest form, the deuteron yield scales as
\begin{equation}
 N_d \propto B_2 N_p N_n ,
\end{equation}
where $B_2$ is the coalescence parameter and $N_p$ and $N_n$ denote
proton and neutron yields in a specified momentum window.

Recent ALICE measurements motivate a closer examination of the
hadronic evolution preceding the formation of the observed deuterons \cite{ALICE2025Nature}.
Resonances and continuum correlation channels can retain and
redistribute baryon number, momentum, and correlations during this
stage.  A finite-lived $\Delta$ population is one particularly simple
example: through $\Delta\to N+\pi$, it provides a delayed source of
nucleons relative to the initial hadronization stage.  Such delayed
injection may become relevant when the surrounding medium has evolved
to conditions under which deuteron destruction is reduced.

This reservoir picture is not intended to replace statistical
hadronization, coalescence, or microscopic transport descriptions.
Nor does the present analysis establish which description gives the
best quantitative account of the ALICE data.  Its purpose is more
restricted: to isolate the dynamical consequence of a finite-lived
intermediate component and to determine how such a component modifies
the time ordering of the formation of a fragile final state.

\subsection{Minimal coarse-grained realization}

At the microscopic level, the hadronic stage of a heavy-ion
collision may be described by an effective hadronic Lagrangian
involving nucleons, pions, and $\Delta$ resonances,
\begin{equation}
\begin{aligned}
{\cal L}={}&
\bar N(i\gamma^\mu\partial_\mu-m_N)N
+\bar\Delta_\mu
 (i\gamma^\nu\partial_\nu-m_\Delta)\Delta^\mu
\\
&+g_{\pi N\Delta}
 \left(\bar\Delta^\mu\partial_\mu\pi\,N
 +{\rm h.c.}\right)+\cdots .
\end{aligned}
\end{equation}
Such a microscopic description can in principle be formulated
in real-time nonequilibrium field theory, for example within
the Schwinger--Keldysh or Kadanoff--Baym framework
\cite{Schwinger1961,Keldysh1965,KB1962}.
After coarse graining over phase space, the resulting kinetic
description may be reduced to evolution equations for selected
densities and correlation channels. A systematic discussion of
this reduction, including the appearance of memory kernels after
finite-lifetime intermediate degrees of freedom are eliminated,
was given in Ref.~\cite{Fuku2026}.

The purpose of the following equations is therefore not to
provide a quantitative transport simulation of a particular
ALICE collision system. In particular, they do not attempt to
implement the complete network of inverse reactions, detailed
balance, realistic medium-dependent rates, or hydrodynamic
expansion and cooling. Instead, they define a minimal
coarse-grained realization with which the dynamical role of an
intermediate reservoir can be isolated.

Let $n_\Delta(t)$, $n_N(t)$, and $n_d(t)$ denote coarse-grained
number densities of $\Delta$ resonances, nucleons, and deuterons.
We consider the representative processes
\begin{align}
\Delta &\rightarrow N+\pi,
\label{eq:delta_decay}\\
N+N &\rightarrow d+X,
\label{eq:NNd}\\
d+X &\rightarrow N+N,
\label{eq:d_break}
\end{align}
where $X$ denotes generic scattering partners in the hadronic
medium.

Additional hadronic correlation channels can be incorporated
in a more complete description. For example, pion--deuteron
continuum correlations, including an effective $d'$-like
correlated channel, may be represented schematically as
\begin{equation}
\pi+d \leftrightarrow d' ,
\label{eq:dprime}
\end{equation}
where $d'$ denotes an effective pion--deuteron correlated
structure. Neither $d'$ nor the $\Delta$ is assumed here to be
a unique microscopic precursor of the final deuteron. They
illustrate how finite-lifetime correlation channels may
contribute to the effective reservoir.

A minimal set of rate equations is
\begin{align}
\dot n_\Delta &=
-\Gamma_\Delta n_\Delta+S_\Delta(t),
\label{eq:ndelta}\\
\dot n_N &=
\Gamma_\Delta n_\Delta
-\langle\sigma_{NN\rightarrow d}v_{\rm rel}\rangle n_N^2
+S_N(t),
\label{eq:nn}\\
\dot n_d &=
\langle\sigma_{NN\rightarrow d}v_{\rm rel}\rangle n_N^2
-\Gamma_d^{\rm br}(T)n_d .
\label{eq:nd}
\end{align}
Here $\Gamma_\Delta$ is an effective reservoir decay rate,
$\Gamma_d^{\rm br}(T)$ is the temperature-dependent deuteron
breakup rate, and $S_\Delta(t)$ and $S_N(t)$ represent effective
source terms associated with the evolving hadronic environment.

The term $\Gamma_\Delta n_\Delta$ should not be interpreted as
an uncorrelated nucleon source. Microscopically, the $\Delta$ is
a correlated $N\pi$ state, and its decay and subsequent
interactions contribute to a network of hadronic correlations.
In a more complete description this information would be
carried by two-body correlation functions and by coupled
channels involving $\Delta N$, $N\pi$, and pion--deuteron
continua. At the present coarse-grained level, their net delayed
contribution is represented by the effective reservoir term
$\Gamma_\Delta n_\Delta$.

Schematically, the corresponding correlated nucleon source may
be written as
\begin{equation}
S_{NN}^{(\Delta)}(t)
\propto
\int d^3x\,C_{NN}^{(\Delta)}({\bf x},t),
\label{eq:SNNDelta}
\end{equation}
where $C_{NN}^{(\Delta)}$ denotes the nucleon--nucleon
correlation associated with the representative
$\Delta$-mediated channel. Eq.~(\ref{eq:ndelta}) is the
simplest realization in which the finite lifetime of the
intermediate component introduces a delay between the
production of the reservoir and its contribution to the final
deuteron channel.

Initially, when the effective source $S_\Delta(t)$ is large and
the medium is hot, $\Gamma_d^{\rm br}(T)$ suppresses the
survival of deuterons. At later times the intermediate
population can remain nonzero while the breakup rate decreases.
The reduced equations then generate a delayed transfer through
the intermediate reservoir. This is the specific dynamical
feature that the numerical examples below are intended to
illustrate.

To illustrate the dynamical consequences of
Eqs.~(\ref{eq:ndelta})--(\ref{eq:nd}), we solve the reduced
cascade equations numerically for representative parameters.
The purpose of this calculation is not to reproduce a specific
heavy-ion collision or to fit the ALICE data, but to display
explicitly the time ordering generated by a finite-lifetime
intermediate reservoir.

For illustration, we take
$\Gamma_\Delta^{-1}=1.5~{\rm fm}/c$, an effective
coalescence-rate parameter
$\lambda\equiv
\langle\sigma_{NN\rightarrow d}v_{\rm rel}\rangle=0.35$,
and a phenomenologically decreasing breakup rate
$\Gamma_d^{\rm br}(t)$. The normalization of the densities is
arbitrary, and the value of $\lambda$ should therefore be
understood as an effective dimensionless parameter introduced
solely to illustrate the generic dynamical behavior of the
reduced equations.

\begin{figure}[t]
\centering
\includegraphics[width=\columnwidth]{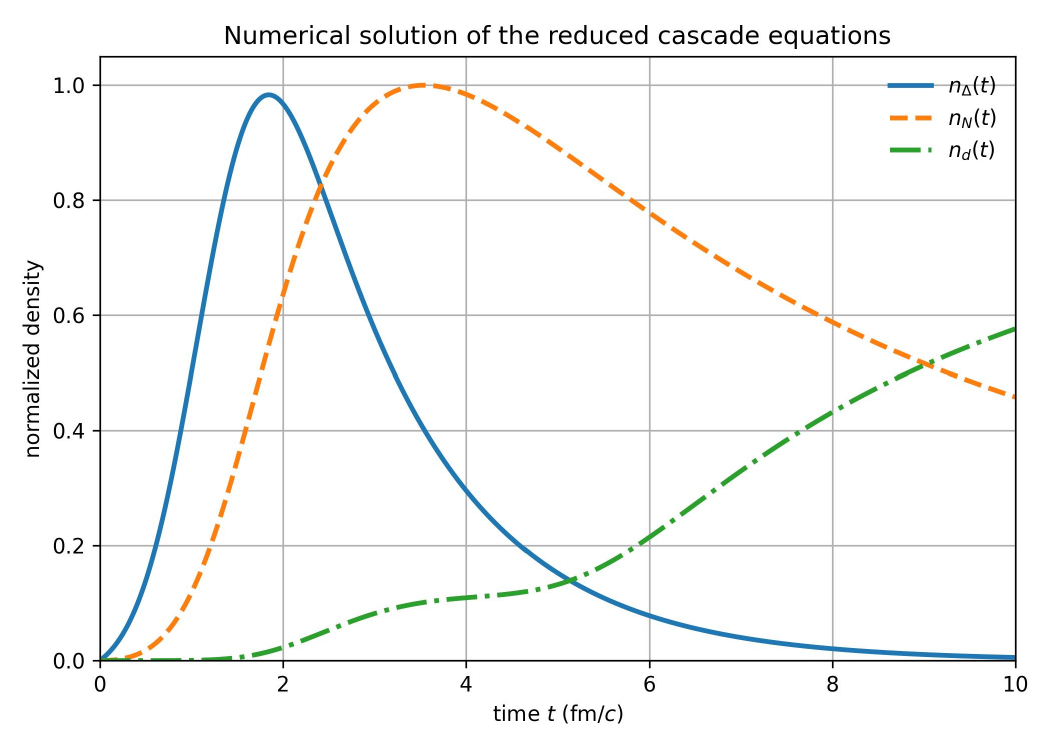}
\caption{
Numerical solution of the reduced cascade equations,
Eqs.~(\ref{eq:ndelta})--(\ref{eq:nd}), for representative
parameters. The intermediate $\Delta$ population is generated
first, the nucleon density receives a delayed contribution
through the finite lifetime of the intermediate reservoir, and
the deuteron abundance grows appreciably only after the
effective breakup rate has decreased. The figure illustrates
the characteristic time ordering of the reduced
reservoir-mediated model; it is not intended as a quantitative
fit to a specific heavy-ion collision system.
}
\label{fig:cascade}
\end{figure}

The resulting evolution is shown in Fig.~\ref{fig:cascade}.
The intermediate $\Delta$ population is generated first and
subsequently feeds the nucleon sector through its finite-lifetime
decay. The deuteron population grows appreciably only at later
times, when the effective breakup rate has decreased. Thus the
numerical solution illustrates the characteristic temporal
ordering generated by the reservoir-mediated cascade,
\[
{\rm reservoir}
\;\longrightarrow\;
{\rm nucleons}
\;\longrightarrow\;
{\rm deuteron}.
\]
This ordering is a property of the reduced model and should not
be interpreted as a quantitative demonstration that a unique
$\Delta$-mediated mechanism describes the ALICE data.

\begin{figure}[t]
\centering
\includegraphics[width=\columnwidth]{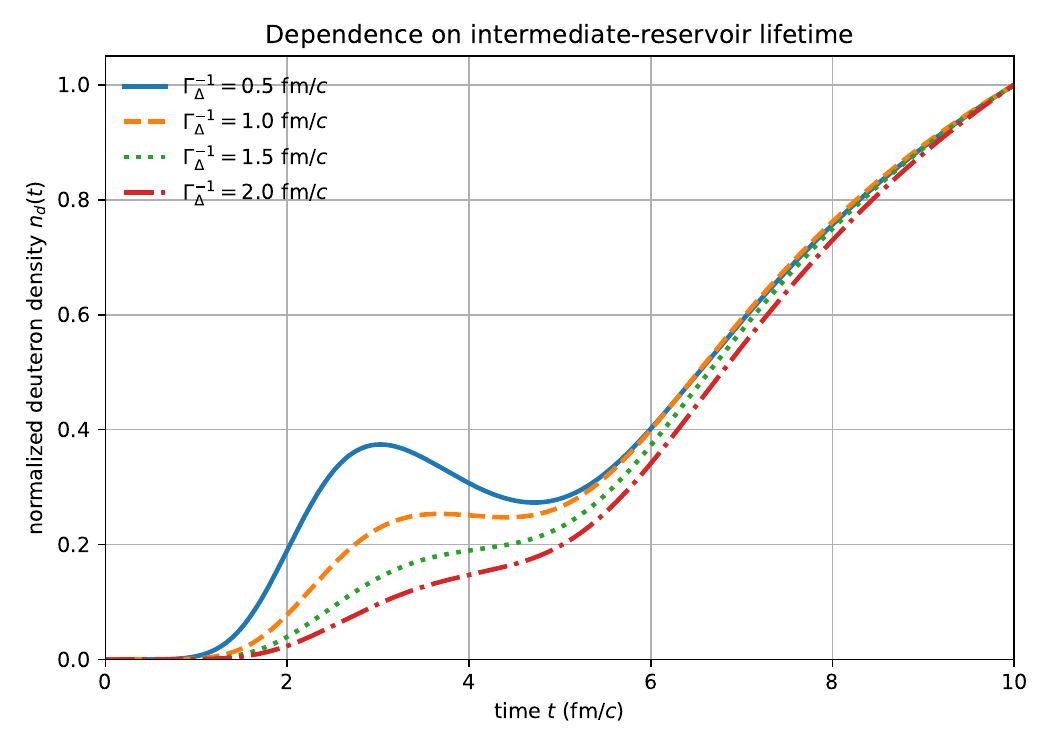}
\caption{
Dependence of the deuteron evolution on the lifetime of the
intermediate reservoir in the reduced cascade model.
Increasing $\Gamma_\Delta^{-1}$ delays the transfer from the
intermediate $\Delta$ population to the nucleon sector and
consequently shifts the buildup of the deuteron component to
later times. The curves illustrate the sensitivity of the
reduced dynamics to the reservoir lifetime; they are not
intended as quantitative predictions for a specific heavy-ion
collision system.
}
\label{fig:lifetime}
\end{figure}

A further consequence of the reservoir picture is that the
formation history depends on the lifetime of the intermediate
component. To illustrate this point, we vary
$\Gamma_\Delta^{-1}$ while keeping the remaining parameters of
the reduced model fixed. The resulting deuteron evolution is
shown in Fig.~\ref{fig:lifetime}.

As the reservoir lifetime increases, the transfer to the
nucleon sector is shifted to later times, and the subsequent
buildup of the deuteron component is correspondingly delayed.
Figure~\ref{fig:lifetime} therefore illustrates directly how
the lifetime of a finite-lived intermediate reservoir controls
the temporal response of the final bound state.

More generally, the importance of such retardation effects is
governed by the ratio
\[
\frac{\tau_R}{\tau_{\rm macro}},
\]
where $\tau_R$ is the characteristic reservoir lifetime and
$\tau_{\rm macro}$ denotes the macroscopic evolution time of
the system. When this ratio is small, an approximately local
(Markovian) description is expected to be adequate. When the
two time scales become comparable, the finite lifetime of the
intermediate component can generate appreciable retardation
and history dependence. In the reduced description obtained
after eliminating the reservoir, these effects appear as a
nonlocal memory kernel \cite{Fuku2026}.

The two numerical examples therefore have complementary roles.
Figure~\ref{fig:cascade} illustrates the time ordering generated
by a finite-lived intermediate reservoir, whereas
Fig.~\ref{fig:lifetime} illustrates the sensitivity of that
ordering to the reservoir lifetime. Together they provide the heavy-ion realization of the reservoir-mediated time ordering discussed in Sec.~II,
without assuming that the same microscopic mechanism
must dominate in all collision systems.
\section{Cosmological Bose--Einstein condensation and
collapse-induced structures}

\subsection{Scalar-field dark matter and reduced
two-component description}

We now turn to cosmological BEC of a scalar field $\phi$
with mass $m$ and self-interaction $g$. At high temperatures
$T\gg T_c$, the field may be described predominantly in terms
of incoherent excitations. As the Universe expands and cools,
macroscopic occupation of a coherent mode can become
dynamically favored. In an equilibrium nonrelativistic Bose
gas, the characteristic condensation temperature is
\begin{equation}
T_c \sim \frac{2\pi}{m}
\left(\frac{n}{\zeta(3/2)}\right)^{2/3}.
\end{equation}
In the cosmological setting considered here, this expression
should be regarded only as a characteristic scale rather than
as evidence that the system remains in strict thermal
equilibrium during its evolution.

In the nonrelativistic regime it is convenient to introduce
a macroscopic wave function $\psi$, with number density
$n=|\psi|^2$. Its dynamics is described schematically by the
Gross--Pitaevskii--Poisson system,
\begin{equation}
i\partial_t\psi =
-\frac{\nabla^2}{2m}\psi
+g|\psi|^2\psi+\Phi\psi,
\qquad
\nabla^2\Phi=4\pi Gm|\psi|^2 .
\end{equation}
Nonlinear self-interaction and gravitational dynamics can then
generate strongly inhomogeneous evolution, including collapse,
re-expansion, and localized structures.

For comparison with the reduced heavy-ion description, it is
useful first to introduce a coarse-grained two-component
description in terms of a coherent condensate fraction
$n_0(t)$ and an excited fraction $n_{\rm ex}(t)$. A minimal
set of effective rate equations is
\begin{align}
\dot n_0 &=
\Gamma_{\rm in}(T)n_{\rm ex}^2
-\Gamma_{\rm out}(T)n_0,
\label{eq:n0}\\
\dot n_{\rm ex} &=
-\Gamma_{\rm in}(T)n_{\rm ex}^2
+\Gamma_{\rm out}(T)n_0
+S_{\rm ex}(t),
\label{eq:nex}
\end{align}
where $\Gamma_{\rm in}(T)$ parametrizes effective transfer
toward the coherent component, $\Gamma_{\rm out}(T)$ describes
the net depletion or redistribution of that component by
nonlinear processes, and $S_{\rm ex}(t)$ includes cosmological
dilution and possible sources. When additional sources are
negligible, the leading dilution contribution may be written
schematically as
$S_{\rm ex}(t)\simeq -3Hn_{\rm ex}$.

The quadratic dependence on $n_{\rm ex}$ in
Eq.~(\ref{eq:n0}) is a minimal parametrization appropriate
when the effective transfer is dominated by two-body
interactions among excited modes. The rate is then
proportional to the number of available excited pairs.

Eqs.~(\ref{eq:n0}) and (\ref{eq:nex}) are deliberately
written as a reduced two-component description. In particular,
the transient localized structures generated during nonlinear
collapse do not appear as an independent variable. Their
effect is contained implicitly in the effective transfer and
depletion rates. As in the heavy-ion example, eliminating a
finite-lived intermediate component can convert a more complete
multi-component dynamics into a reduced description with
effective rates and, in general, memory dependence
\cite{Fuku2026}. The underlying intermediate component becomes
explicit in the original FMT formulation discussed below.

\subsection{FMT three-component formulation and
collapse-induced reservoir}

The original FMT framework \cite{FMT2008} is formulated in
terms of three coarse-grained energy-density components: a
boson-gas component $\rho_g$, a coherent component $\rho_\phi$,
and a localized collapse-product component $\rho_l$. In a
spatially flat FRW background one writes
\begin{equation}
H^2=\left(\frac{\dot a}{a}\right)^2
=\frac{8\pi G}{3c^2}
(\rho_g+\rho_\phi+\rho_l),
\label{eq:FMT_H}
\end{equation}
\begin{equation}
\dot\rho_g=-3H\rho_g-\Gamma\rho_g,
\label{eq:FMT_rg}
\end{equation}
\begin{equation}
\dot\rho_\phi=
-6H(\rho_\phi-V)
+\Gamma\rho_g-\Gamma'\rho_\phi,
\label{eq:FMT_rphi}
\end{equation}
\begin{equation}
\dot\rho_l=
-3H\rho_l+\Gamma'\rho_\phi,
\label{eq:FMT_rl}
\end{equation}
where $\Gamma$ is an effective transfer rate from the boson-gas
component to the coherent component, and $\Gamma'$ describes
transfer from the coherent component to localized structures.
The latter arise from nonlinear collapse driven by
gravitational and self-interaction effects.

The three-component formulation makes explicit a feature that
is hidden in the reduced two-component equations: the coherent
component is dynamically coupled to a distinct localized sector
generated by nonlinear collapse.  In the FMT equations, the
dominant coarse-grained transfer is organized as
\[
 \rho_g \longrightarrow \rho_\phi,
 \qquad
 \rho_\phi \longrightarrow \rho_l ,
\]
where the first process describes transfer from the boson-gas
component to the coherent component, while the second describes
the formation of localized structures through nonlinear collapse.
Thus $\rho_l$ should not be interpreted as a microscopic
precursor through which all particles must pass before entering
the condensate.

Nevertheless, the localized component introduces an additional
nonequilibrium time scale into the evolution of the coherent
sector.  Collapse, localization, and subsequent redistribution
can interrupt a smooth monotonic approach toward condensation
and make the coarse-grained evolution dependent on its previous
history.  In this sense, the localized sector plays the role of
a transient nonequilibrium reservoir coupled to the condensate,
although its position in the dynamical network differs from
that of the $\Delta$ reservoir in the heavy-ion example.

Eqs.~(\ref{eq:n0}) and (\ref{eq:nex}) should consequently
not be regarded as a direct rewriting of
Eqs.~(\ref{eq:FMT_rg})--(\ref{eq:FMT_rl}).  They provide a
minimal two-component parametrization of transfer between
excited and coherent components, whereas the original FMT
equations retain explicitly the additional localized sector.
The relation between the two descriptions is therefore one of
coarse graining rather than a one-to-one identification of
their variables.

The comparison with the heavy-ion system should accordingly be
made at the level of dynamical organization.  In both cases a
fragile or coherent final component evolves in the presence of
finite-lived or transient nonequilibrium degrees of freedom
that introduce additional time scales and can produce delayed
or history-dependent behavior.  The topology of the microscopic
transfer network, however, need not be identical.

\begin{figure}[t]
\centering
\includegraphics[width=\columnwidth]{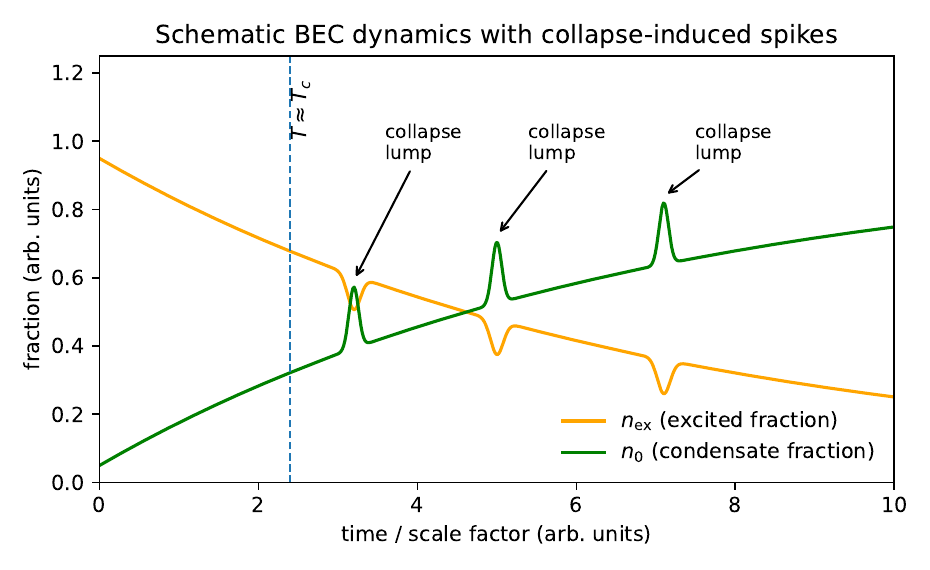}
\caption{
Schematic illustration of the nonequilibrium dynamics of
cosmological Bose--Einstein condensation motivated by the FMT
framework \cite{FMT2008}.  The excited component
$n_{\rm ex}$ decreases while the coherent component $n_0$
develops as the system evolves through the condensation regime.
Nonlinear collapse and re-expansion generate transient localized
structures, indicated schematically by the spikes.  These
structures represent an additional nonequilibrium sector coupled
to the coherent component and can interrupt a smooth monotonic
evolution.  The figure is intended to illustrate the temporal
organization of the FMT scenario rather than a universal
quantitative evolution.
}
\label{fig:BEC}
\end{figure}

The corresponding cosmological evolution is illustrated
schematically in Fig.~\ref{fig:BEC}.

\section{Common nonequilibrium structure and its limitations}

The two systems discussed in Secs.~II and III involve very
different microscopic physics.  Light-nuclei formation takes
place in a short-lived hadronic medium and is governed by
strong-interaction reaction and correlation channels, whereas
cosmological BEC evolves on vastly longer time scales under
self-interaction, gravity, and cosmic expansion.  The
correspondence proposed here therefore does not require, and
should not be interpreted as, an identification of their
microscopic reaction networks.

\begin{figure}[t]
\centering
\includegraphics[width=\columnwidth]{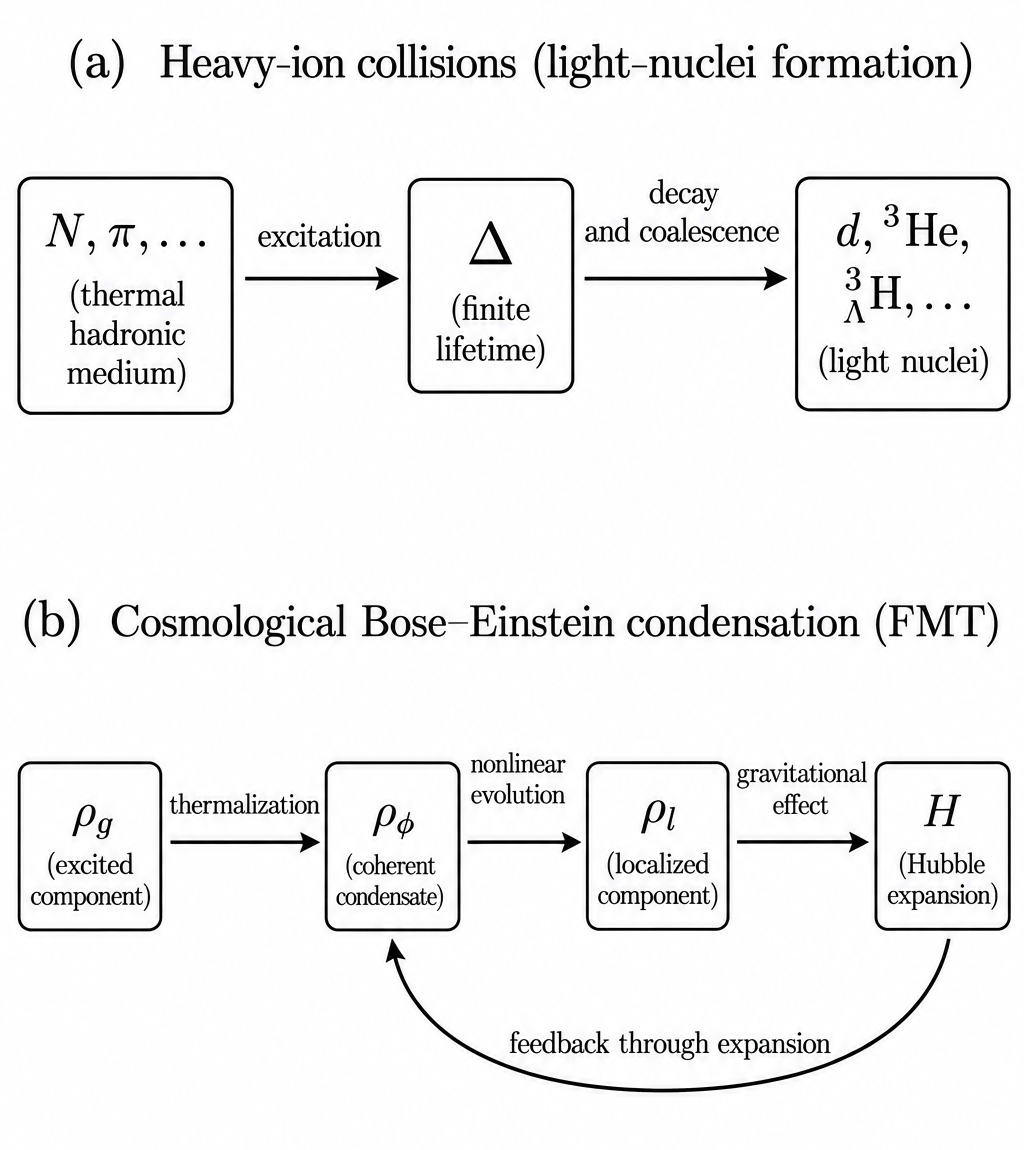}
\caption{
Schematic comparison of the time ordering and coupling
structure in (a) light-nuclei formation in heavy-ion
collisions and (b) cosmological Bose--Einstein condensation
in the FMT scenario.
(a) Thermal hadronic matter excites intermediate
finite-lifetime components, represented here by the
$\Delta$ resonance. Their subsequent decay and the
resulting hadronic evolution feed the formation of light
nuclei. Eliminating the finite-lived intermediate sector
generates a causal memory term in the reduced dynamics.
(b) In the cosmological system, the excited component
$\rho_g$ evolves toward the coherent condensate
$\rho_\phi$, whose nonlinear evolution generates the
localized component $\rho_l$. The localized component
contributes to the Hubble expansion rate $H$, which in
turn feeds back into the evolution of $\rho_\phi$.
The resulting loop
$\rho_\phi\to\rho_l\to H\to\rho_\phi$
is the origin of the causal memory that appears after the
localized sector is eliminated.
The two diagrams therefore do not represent identical
microscopic cascade topologies; rather, they illustrate
two different physical realizations of the same reduced
dynamical structure in which elimination of an additional
coupled sector leaves a history-dependent contribution.
}
\label{fig:mapping}
\end{figure}

Figure~\ref{fig:mapping} summarizes both the similarity and
the difference.  In the heavy-ion example, the reduced
dynamical sequence considered in Sec.~II is
\[
 {\rm hadronic\ sector}
 \longrightarrow
 \Delta/{\rm correlations}
 \longrightarrow
 {\rm light\ nuclei}.
\]
The finite lifetime of the intermediate hadronic component
introduces a delay before correlated nucleons become available
under conditions in which a fragile deuteron can survive.

The cosmological system has a different dynamical topology.
In the FMT description, the corresponding coarse-grained
evolution is organized as
\[
 \rho_g \longrightarrow \rho_\phi,
 \qquad
 \rho_\phi \longrightarrow \rho_l ,
\]
or, in the notation used for the reduced description,
schematically as
\[
 n_{\rm ex}
 \longrightarrow
 n_0
 \longrightarrow
 {\rm localized\ structures}.
\]
Here the localized structures are generated by nonlinear
collapse of the coherent component.  They are therefore not
the analogue of a microscopic precursor through which the
system must pass before condensation.  Rather, they constitute
an additional transient nonequilibrium sector coupled to the
coherent component.

\subsection{What is universal}

The comparison between light-nuclei formation and
cosmological Bose--Einstein condensation should not be
understood as a one-to-one correspondence between their
microscopic reaction networks.  In particular, the
$\Delta$ resonance in the hadronic system and the localized
component $\rho_l$ in the FMT system occupy different
positions in their respective dynamical networks.  The
common structure instead appears at the level of reduced
nonequilibrium dynamics.

A general feature of a multicomponent nonequilibrium system
is that the elimination of an auxiliary dynamical sector can
leave a causal dependence on the previous evolution of the
retained variables \cite{BP2007}.  Schematically, if a retained variable
$X$ is dynamically coupled, directly or indirectly, to an
additional variable $R$,
\begin{equation}
 \dot X=A X+B R,\qquad
 \dot R=C X+D R ,
 \label{eq:general_reduced}
\end{equation}
the formal solution for $R$ is
\begin{equation}
 R(t)=G_R(t,t_0)R(t_0)
 +\int_{t_0}^{t}dt'\,
 G_R(t,t')C(t')X(t'),
 \label{eq:Rformal}
\end{equation}
where
\begin{equation}
 G_R(t,t')
 =
 \exp\left[
 \int_{t'}^{t}ds\,D(s)
 \right]
\end{equation}
for scalar time-dependent coefficients.  Substitution into
the equation for $X$ gives
\begin{equation}
 \dot X(t)
 =
 A(t)X(t)
 +B(t)G_R(t,t_0)R(t_0)
 +\int_{t_0}^{t}dt'\,
 {\cal K}(t,t')X(t'),
 \label{eq:memory_general}
\end{equation}
with
\begin{equation}
 {\cal K}(t,t')
 =
 B(t)G_R(t,t')C(t').
 \label{eq:memory_kernel}
\end{equation}
Thus, when the response time of the eliminated sector cannot
be neglected, its reduction generally produces a
history-dependent effective dynamics.  This reduction has a direct connection with the Schwinger--Keldysh formulation developed in
Ref.~\cite{Fuku2026}.  More generally, Schwinger--Keldysh
approaches provide a natural framework for describing
non-Markovian reduced dynamics beyond the instantaneous
Markovian limit~\cite{RO2026}. There, the effective evolution of a
retained degree of freedom was shown to acquire a nonlocal
term of the generic form
\begin{equation}
 \dot Y(t)
 =
 F[Y(t)]
 +
 \int_{t_0}^{t}dt'\,
 K(t,t')\,{\cal S}[Y(t')],
 \label{eq:SK_memory}
\end{equation}
when finite-time correlations associated with eliminated
degrees of freedom are retained.  Eq.~(\ref{eq:SK_memory}) is the central universal structure considered in this work. The heavy-ion and FMT systems constitute two microscopically distinct realizations of this same reduced nonequilibrium equation. The usual local rate
equation is recovered in the Markovian limit, in which the
correlation or memory time is short compared with the
macroscopic evolution time.
The present analysis provides two concrete realizations of
this general reduced-dynamical structure.  In the heavy-ion
system, elimination of the finite-lived $\Delta$ component
produces a retarded contribution with a characteristic scale
$\Gamma_\Delta^{-1}$.  In the FMT system, the mechanism is
different: elimination of the localized component $\rho_l$,
together with its contribution to the Friedmann equation,
produces a history dependence through the feedback
\begin{equation}
 \rho_\phi\longrightarrow\rho_l
 \longrightarrow H\longrightarrow\rho_\phi .
 \label{eq:sequence}
\end{equation}
For comparison, the corresponding heavy-ion dynamics may be
schematically organized as
\begin{equation}
\rho_{\rm had}
\;\longrightarrow\;
\Delta + N
\;\longrightarrow\;
N + N + \pi
\;\longrightarrow\;
d + \pi ,
\label{eq:HI_sequence}
\end{equation}
where $\rho_{\rm had}$ denotes the initially excited hadronic sector.
The two sequences are not microscopically identical: the heavy-ion
case describes a delayed transfer through a finite-lived resonance
sector, whereas Eq.~(\ref{eq:sequence}) represents a gravitational feedback loop.
Their common feature emerges only after the additional dynamical
sector is eliminated, leading to a causal memory in the reduced
dynamics.
Thus, Ref.~\cite{Fuku2026} supplies the general
Schwinger--Keldysh basis for memory in reduced nonequilibrium
dynamics, while the present work identifies how such a
structure is realized in two physically distinct systems.
The corresponding kernels need not have the same functional
form.

In the heavy-ion example this structure is particularly
transparent.  From Eq.~(\ref{eq:ndelta}), the resonance
component obeys
\begin{equation}
 \dot n_\Delta
 =
 -\Gamma_\Delta n_\Delta+S_\Delta(t),
\end{equation}
and hence
\begin{equation}
 n_\Delta(t)
 =
 e^{-\Gamma_\Delta(t-t_0)}n_\Delta(t_0)
 +
 \int_{t_0}^{t}dt'\,
 e^{-\Gamma_\Delta(t-t')}S_\Delta(t').
 \label{ndelta2}
\end{equation}
Substituting Eq.~(\ref{ndelta2}) into the nucleon
equation (\ref{eq:nn}) gives
\begin{align}
 \dot n_N(t)
 ={}&
 -\lambda n_N^2(t)+S_N(t)
 +\Gamma_\Delta e^{-\Gamma_\Delta(t-t_0)}
 n_\Delta(t_0)
 \nonumber\\
 &+
 \int_{t_0}^{t}dt'\,
 {\cal K}_{\rm HI}(t,t')S_\Delta(t'),
 \label{eq:HI_memory}
\end{align}
where
\begin{equation}
 {\cal K}_{\rm HI}(t,t')
 =
 \Gamma_\Delta
 e^{-\Gamma_\Delta(t-t')}.
 \label{eq:HI_kernel}
\end{equation}
Here
$\lambda\equiv\langle\sigma_{NN\to d}v_{\rm rel}\rangle$.
Thus, after elimination of the intermediate $\Delta$
component, the retained nucleon dynamics acquires explicitly
the same causal reduced structure as Eq.~(\ref{eq:SK_memory}),
with a memory time
\begin{equation}
 \tau_\Delta=\Gamma_\Delta^{-1}.
\end{equation}
In the limit $\tau_\Delta\ll\tau_{\rm macro}$ the kernel
becomes sharply localized in time, whereas for
$\tau_\Delta\sim\tau_{\rm macro}$ the finite temporal width
of the kernel must be retained.  The relevant distinction is therefore
between the limit
\begin{equation}
 \tau_\Delta\ll\tau_{\rm macro},
\end{equation}
in which the intermediate sector may be eliminated
approximately instantaneously, and the regime in which the
two time scales become comparable and retardation must be
retained.

A related, but microscopically different, structure is present
in the FMT system.  The localized component satisfies
Eq.~(\ref{eq:FMT_rl}).
Using $H=\dot a/a$, its formal solution is
\begin{equation}
 \rho_l(t)
 =
 \left[\frac{a(t_0)}{a(t)}\right]^3\rho_l(t_0)
 +
 \int_{t_0}^{t}dt'\,
 \left[\frac{a(t')}{a(t)}\right]^3
 \Gamma'(t')\rho_\phi(t').
 \label{eq:rhol_formal}
\end{equation}
The localized component therefore contains information about
the previous evolution of the coherent component.  This fact
alone would not affect $\rho_\phi$ if the expansion rate were
externally prescribed.  In the full FMT system, however,
$\rho_l$ contributes to the Hubble rate through the Friedmann
equation, Eq.~(\ref{eq:FMT_H}).
Since $H$ in turn enters Eq.~(\ref{eq:FMT_rphi}), the full
system contains the indirect feedback loop
\begin{equation}
 \rho_\phi\longrightarrow\rho_l
 \longrightarrow H\longrightarrow\rho_\phi .
 \label{eq:FMT_feedback}
\end{equation}
Eliminating $\rho_l$ while retaining the Friedmann constraint
therefore converts the local three-component description into
a history-dependent reduced dynamics for the retained
variables.

This can be seen explicitly by linearizing about a background
solution,
\begin{equation}
 \rho_i=\bar\rho_i+\delta\rho_i,\qquad
 H=\bar H+\delta H .
\end{equation}
Writing
\begin{equation}
 C_G\equiv\frac{8\pi G}{3c^2},
\end{equation}
the linearized Friedmann equation gives
\begin{equation}
 \delta H=
 \frac{C_G}{2\bar H}
 \left(
 \delta\rho_g+\delta\rho_\phi+\delta\rho_l
 \right).
 \label{eq:deltaH}
\end{equation}
Consequently, the coherent component contains a term
\begin{equation}
 \delta\dot\rho_\phi
 =
 \cdots+B_\phi(t)\delta\rho_l,
\end{equation}
where
\begin{equation}
 B_\phi(t)
 =
 -\frac{3C_G}{\bar H(t)}
 \left[\bar\rho_\phi(t)-\bar V(t)\right].
 \label{eq:Bphi}
\end{equation}
Similarly, the linearized localized component can be written
as
\begin{equation}
 \delta\dot\rho_l
 =
 A_l(t)\delta\rho_l
 +C_l(t)\delta\rho_\phi+\cdots .
 \label{delta_rhol}
\end{equation}
Eliminating $\delta\rho_l$ then gives
\begin{equation}
 \delta\dot\rho_\phi(t)
 =
 A_\phi(t)\delta\rho_\phi(t)
 +
 \int_{t_0}^{t}dt'\,
 {\cal K}_{\rm FMT}(t,t')
 \delta\rho_\phi(t')
 +\cdots ,
 \label{eq:FMT_memory}
\end{equation}
with
\begin{equation}
 {\cal K}_{\rm FMT}(t,t')
 =
 B_\phi(t)G_l(t,t')C_l(t').
 \label{eq:FMT_kernel}
\end{equation}
For a subdominant localized background,
$G_l(t,t')$ reduces approximately to the cosmological
dilution factor
\begin{equation}
 G_l(t,t')
 \simeq
 \left[\frac{\bar a(t')}{\bar a(t)}\right]^3 .
\end{equation}

The two memory structures are therefore not identical.  In
the hadronic problem the retardation is controlled directly
by the finite lifetime of an intermediate excitation, whereas
in the FMT system the history dependence is mediated
indirectly through the gravitational expansion.  The detailed
kernels, time scales, and microscopic mechanisms are
system-dependent.  What is common is the reduced-dynamical
principle: eliminating a dynamically coupled sector with a
non-negligible response time can leave a causal memory of that
sector in the effective evolution of the retained variables.
It is in this restricted sense that universality is used in
the present work.

\subsection{System-dependent realizations of the universal structure}

The significance of the common structure identified above
lies precisely in the fact that its microscopic realizations
are different.  Universality does not require the underlying
degrees of freedom, interactions, or characteristic time
scales to be identical.  Rather, it refers to a dynamical
structure that survives these microscopic differences after
the appropriate degrees of freedom have been coarse grained.

In the heavy-ion system, the additional dynamical sector is
represented in the minimal model by a finite-lived
$\Delta$ component.  Its finite response time is set directly
by $\Gamma_\Delta^{-1}$, and eliminating this component
generates an explicitly retarded contribution to the
subsequent nucleon and correlation dynamics.  In the simplest
case, the corresponding memory has the characteristic form
\begin{equation}
 {\cal K}_{\rm HI}(t,t')
 \sim
 e^{-\Gamma_\Delta(t-t')}.
 \label{eq:KHI}
\end{equation}
The physical origin of the memory is therefore the finite
lifetime of an intermediate hadronic excitation.

The cosmological realization is microscopically very
different.  The FMT dynamics is organized as
\begin{equation}
 \rho_g\longrightarrow\rho_\phi\longrightarrow\rho_l ,
 \label{eq:FMTsequence}
\end{equation}
so that the localized component $\rho_l$ is generated from
the coherent component rather than serving as a microscopic
precursor to it.  Nevertheless, the Friedmann equation closes
the dynamical loop,
\begin{equation}
 \rho_\phi\longrightarrow\rho_l
 \longrightarrow H\longrightarrow\rho_\phi ,
 \label{eq:FMTloop}
\end{equation}
and the elimination of $\rho_l$ therefore leaves information
about its previous evolution in the reduced coherent
dynamics.  As derived above, the linearized memory kernel has
the form
\begin{equation}
 {\cal K}_{\rm FMT}(t,t')
 =
 B_\phi(t)G_l(t,t')C_l(t').
 \label{eq:KFMT}
\end{equation}
Here the memory is mediated by gravitational expansion rather
than by the direct decay of a finite-lived resonance.

Eqs.~(\ref{eq:KHI}) and (\ref{eq:KFMT}) are not expected
to have the same functional form.  Their difference is not an
obstruction to the comparison; it identifies two distinct
physical realizations of the same reduced-dynamical
principle.  In both cases the full dynamics contains an
additional sector that cannot, in general, be removed
instantaneously.  Once that sector is eliminated, its
previous evolution can remain encoded in the dynamics of the
retained variables through a causal memory term.

This observation also clarifies the relation to the general
real-time framework of Ref.~\cite{Fuku2026}.  The
Schwinger--Keldysh reduction discussed there provides the
general mechanism by which unresolved finite-time dynamics
can appear as a nonlocal contribution to an effective
evolution equation.  The present examples show how this
principle is realized in two systems whose microscopic
physics is otherwise widely separated: directly through a
finite-lived hadronic component in one case, and indirectly
through gravitational feedback in the other.

The comparison therefore concerns neither an identification
$\Delta\equiv\rho_l$ nor an identical ordering of microscopic
processes.  At the level of physical interpretation,
$n_{\rm ex}$ in the reduced cosmological description is most
closely associated with the boson-gas sector $\rho_g$, while
$n_0$ represents the coherent sector $\rho_\phi$.  The
localized component $\rho_l$ has no independent counterpart
in Eqs.~(\ref{eq:n0}) and (\ref{eq:nex}); its influence is
contained in the coarse-grained dynamics.  This is precisely
why the comparison is made at the level of reduced dynamics
rather than by a term-by-term identification of the original
equations.

The quantitative importance of the resulting memory remains
system dependent.  For example, the gravitational feedback
in the FMT system becomes weak when the localized component
is dynamically negligible or when the coefficient
$B_\phi$ in Eq.~(\ref{eq:Bphi}) is suppressed.  Likewise,
the heavy-ion dynamics approaches a local description when
$\Gamma_\Delta^{-1}$ is short compared with the macroscopic
evolution time.  These limits determine the strength of a
particular realization; they do not alter the common
reduction principle.

\subsection{Implications of the universal structure}

The preceding analysis suggests a general criterion for
deciding when a reduced nonequilibrium description can be
approximated by instantaneous rates and when information
about the previous evolution must be retained.  The relevant
question is not the microscopic identity of the eliminated
degrees of freedom, but whether their response time is short
compared with the characteristic evolution time of the
retained sector.

Consider schematically an additional dynamical sector $R$
coupled to a retained variable $X$.  If the characteristic
response time $\tau_R$ of $R$ satisfies
\begin{equation}
 \tau_R \ll \tau_{\rm macro},
 \label{eq:Markovcriterion}
\end{equation}
the eliminated sector follows the retained variables
adiabatically and its effect can, to leading order, be
absorbed into instantaneous effective rates.  This is the
Markovian limit of the reduced dynamics.  When
\begin{equation}
 \tau_R \sim \tau_{\rm macro},
 \label{eq:memorycriterion}
\end{equation}
the evolution of the eliminated sector cannot in general be
represented by its instantaneous value.  Its previous
evolution remains encoded in the reduced equation through a
causal memory term.  This is the time-scale criterion
underlying the general real-time reduction discussed in
Ref.~\cite{Fuku2026}.

The heavy-ion example provides a particularly transparent
realization.  The response time of the representative
intermediate sector is
\begin{equation}
 \tau_\Delta \sim \Gamma_\Delta^{-1}.
\end{equation}
Eq.~(\ref{ndelta2}) shows explicitly that the
$\Delta$ population at time $t$ retains information about
its earlier source history.  When $\tau_\Delta$ is much
shorter than the macroscopic evolution time, the exponential
kernel becomes sharply localized and an approximately local
description is recovered.  When the two time scales become
comparable, the finite width of the kernel must be retained
and the reduced nucleon and correlation dynamics becomes
history dependent.

Figures~1 and 2 illustrate this transition at the level of
the reduced heavy-ion model.  The finite response time of the
intermediate sector delays the transfer to the nucleon sector,
and changing $\Gamma_\Delta^{-1}$ changes the subsequent
time evolution of the deuteron component.  The significance
of these figures is therefore not that they select a unique
microscopic production channel, but that they display directly
how an unresolved dynamical time scale survives in the
evolution of a retained observable.

The FMT system realizes the same principle in a less direct
and therefore particularly instructive way.  The localized
component is generated from the coherent sector,
\begin{equation}
 \rho_\phi\longrightarrow\rho_l,
\end{equation}
but the Friedmann equation closes the feedback through
\begin{equation}
 \rho_l\longrightarrow H\longrightarrow\rho_\phi .
\end{equation}
Consequently, the history stored in $\rho_l$ is transmitted
back to the coherent dynamics through the expansion rate.
Eq.~(\ref{eq:rhol_formal}) makes the stored history
explicit, while Eqs.~(\ref{eq:FMT_memory}) and
(\ref{eq:FMT_kernel}) show that eliminating the localized
sector produces a causal memory kernel in the linearized
reduced dynamics.

This result is important because the cosmological memory does
not arise from a finite-lived resonance analogous to the
$\Delta$.  It is generated by a different microscopic
mechanism and is communicated to the retained sector through
gravitational feedback.  Nevertheless, after reduction both
systems take the generic form
\begin{equation}
 \dot X(t)
 =
 F[X(t)]
 +
 \int_{t_0}^{t}dt'\,
 {\cal K}(t,t')\,{\cal S}[X(t')]
 +\cdots ,
 \label{eq:universal_reduced}
\end{equation}
where the detailed form of ${\cal K}$ is determined by the
physical realization.  The common object is therefore not a
particular intermediate particle or a particular cascade
topology, but the causal reduced dynamics that remains after
a dynamically relevant sector has been eliminated.

This perspective also clarifies why very different
microscopic systems can belong to the same reduced-dynamical
class.  The hadronic and cosmological examples differ in
their constituents, interactions, absolute time scales, and
mechanisms of feedback.  Such differences determine the
specific kernel, but they do not remove the common operation
of dynamical reduction.  Indeed, the persistence of the same
causal structure after these microscopic details have been
changed is precisely what makes the comparison nontrivial.

The connection with Ref.~\cite{Fuku2026} can therefore be
stated more sharply.  Ref.~\cite{Fuku2026} provides the
general Schwinger--Keldysh basis for the emergence of
nonlocal terms when finite-time dynamics is retained under
coarse graining.  The present analysis identifies two
physically distinct realizations of this principle and shows
explicitly how their corresponding memory structures arise:
through the lifetime of a hadronic intermediate sector in
one case and through the
$\rho_\phi\rightarrow\rho_l\rightarrow H\rightarrow\rho_\phi$
feedback loop in the other.

A broader implication is that an equilibrium-like final
state does not uniquely determine the dynamical path by
which it was reached.  Coarse graining can hide intermediate
or auxiliary dynamical sectors even when final observables
admit a successful statistical or equilibrium
parametrization.  Conversely, the existence of such hidden
nonequilibrium dynamics need not prevent equilibrium-like
regularities from emerging in final observables.  What is
lost in an instantaneous description is the temporal
information carried by the eliminated degrees of freedom.

The resulting diagnostic principle is therefore general:
one should ask whether the degrees of freedom removed in
constructing an effective description respond
instantaneously on the time scale of the retained dynamics.
If they do, a local effective description is appropriate.
If they do not, their history survives as memory.  This
criterion is independent of whether the underlying system is
hadronic, cosmological, or belongs to another microscopic
domain, and it is this independence that gives the reduced
nonequilibrium structure its universal character.
\subsection{Predictions beyond the two realizations}
\label{sec:predictions}

The reduced-dynamical formulation developed above is not
restricted to the two examples considered in this work.
Once the universal content is identified with the causal
structure of Eq.~(\ref{eq:SK_memory}), rather than with a particular
microscopic cascade, it leads to predictions for other
nonequilibrium systems containing dynamically relevant
degrees of freedom that are not retained in an effective
description.

A particularly direct consequence concerns systems with
more than one intermediate dynamical sector.  To see this,
consider a set of additional variables collected into a
vector ${\bf R}$, coupled to a retained variable $Y$,
\begin{align}
 \dot Y(t)
 &=
 F[Y(t)]+{\bf B}(t){\bf R}(t),
 \label{eq:multiY}
 \\
 \dot{\bf R}(t)
 &=
 -{\bf \Gamma}(t){\bf R}(t)
 +{\bf C}(t){\cal S}[Y(t)] .
 \label{eq:multiR}
\end{align}
For a time-independent matrix ${\bf \Gamma}$, the formal
solution of Eq.~(\ref{eq:multiR}) is
\begin{align}
 {\bf R}(t)
 &=
 e^{-{\bf \Gamma}(t-t_0)}{\bf R}(t_0)
 \nonumber\\
 &\quad+
 \int_{t_0}^{t}dt'\,
 e^{-{\bf \Gamma}(t-t')}
 {\bf C}(t'){\cal S}[Y(t')] .
 \label{eq:multiRsol}
\end{align}
Substitution into Eq.~(\ref{eq:multiY}) gives the memory
kernel
\begin{equation}
 \boxed{
 {\cal K}(t,t')
 =
 {\bf B}(t)
 e^{-{\bf \Gamma}(t-t')}
 {\bf C}(t')
 } .
 \label{eq:multikernel}
\end{equation}
Thus the spectrum of relaxation modes of the eliminated
sector is directly encoded in the temporal structure of
the effective kernel.

If ${\bf \Gamma}$ possesses several relaxation eigenvalues
$\gamma_i$, the kernel contains, in the simplest
diagonalizable case, several characteristic time scales,
\begin{equation}
 {\cal K}(t,t')
 \sim
 \sum_i A_i(t,t')
 e^{-\gamma_i(t-t')}.
 \label{eq:multiscale}
\end{equation}
The single exponential kernel of the minimal heavy-ion
example is therefore only the simplest member of a broader
class.  A system whose formation dynamics involves several
dynamically relevant intermediate channels is predicted to
retain a correspondingly richer memory structure after
those channels are eliminated.
This result suggests a direct way of testing the
multi-scale prediction in composite-particle formation.
One should consider systems in which the final bound state
can be reached through more than one dynamically relevant
formation stage.  Light nuclei and hypernuclei provide a
natural testing ground for this purpose, since increasing
the complexity of the composite state can introduce
additional intermediate formation channels.  A particularly
simple example beyond the deuteron is the hypertriton
${}^{3}_{\Lambda}{\rm H}$. For the hypertriton, the two representative formation
pathways may be written schematically as \cite{ZhangKo2018}
\begin{align}
p+n+\Lambda
&\longrightarrow {}^3_{\Lambda}{\rm H}
\qquad \text{(direct)}, \nonumber\\
p+n
&\longrightarrow d
\longrightarrow {}^3_{\Lambda}{\rm H}
\qquad \text{(sequential)}.
\label{hyper}
\end{align}
Here the deuteron in the sequential pathway constitutes an
additional intermediate dynamical sector that is absent from the
direct channel.  The existence of two formation pathways does not,
by itself, imply two independent exponential relaxation modes.
Rather, the minimal consequence of the sequential channel is that
eliminating the intermediate deuteron sector generates a retarded
contribution to the reduced hypertriton dynamics.

To make this point explicit, let
$n_H\equiv n_{{}^3_{\Lambda}{\rm H}}$ and write a minimal
coarse-grained description as
\begin{align}
 \dot n_d(t)
 &=
 S_d(t)-\Gamma_d^{\rm eff}(t)n_d(t),
 \label{hyper_d}\\
 \dot n_H(t)
 &=
 S_{\rm dir}(t)
 +\lambda_{d\Lambda}(t)n_\Lambda(t)n_d(t)
 -\Gamma_H^{\rm br}(t)n_H(t),
 \label{hyper_H}
\end{align}
where
\begin{equation}
 S_{\rm dir}(t)
 =
 \lambda_{pn\Lambda}(t)
 n_p(t)n_n(t)n_\Lambda(t)
 \label{hyper_direction}
\end{equation}
represents the direct $p+n+\Lambda$ formation channel.
Here $\Gamma_d^{\rm eff}$ denotes an effective rate governing
the evolution of the intermediate deuteron sector, including
processes that remove or redistribute deuterons in the evolving
hadronic medium.

The formal solution of Eq.~(\ref{hyper_d}) is
\begin{equation}
 n_d(t)
 =
 G_d(t,t_0)n_d(t_0)
 +
 \int_{t_0}^{t}dt'\,
 G_d(t,t')S_d(t'),
 \label{hyper_d_solution}
\end{equation}
where
\begin{equation}
 G_d(t,t')
 =
 \exp\left[
 -\int_{t'}^{t}ds\,
 \Gamma_d^{\rm eff}(s)
 \right].
 \label{hyper_Gd}
\end{equation}
Substitution into Eq.~(\ref{hyper_H}) gives
\begin{align}
 \dot n_H(t)
 ={}&
 S_{\rm dir}(t)
 -\Gamma_H^{\rm br}(t)n_H(t)
 \nonumber\\
 &+
 \lambda_{d\Lambda}(t)n_\Lambda(t)
 G_d(t,t_0)n_d(t_0)
 \nonumber\\
 &+
 \int_{t_0}^{t}dt'\,
 K_{\rm seq}(t,t')S_d(t'),
 \label{hyper_memory}
\end{align}
with
\begin{equation}
 K_{\rm seq}(t,t')
 =
 \lambda_{d\Lambda}(t)n_\Lambda(t)G_d(t,t').
 \label{hyper_kernel}
\end{equation}
Thus the direct and sequential pathways have different roles in
the reduced dynamics.  The direct channel appears as a local
source, whereas elimination of the intermediate deuteron sector
generates an explicitly causal memory contribution.  This is the
hypertriton realization of the general reduction mechanism of
Eqs.~(\ref{eq:multiY})--(\ref{eq:multiscale}).

A useful quantitative benchmark is provided by the coalescence
calculation of Ref.~\cite{ZhangKo2018}.  For central Pb--Pb
collisions at $\sqrt{s_{NN}}=2.76$ TeV, the direct
$p+n+\Lambda$ and sequential $d+\Lambda$ contributions to the
hypertriton yield were found to be approximately
\begin{equation}
 \left.
 \frac{dN_H}{dy}
 \right|_{\rm dir}
 \simeq 6.7\times10^{-5},
 \qquad
 \left.
 \frac{dN_H}{dy}
 \right|_{\rm seq}
 \simeq 7.9\times10^{-5}.
 \label{hyper_yields}
\end{equation}
The corresponding initial ratio is therefore
\begin{equation}
 R_{\rm seq/dir}(0)
 \equiv
 \frac{N_{\rm seq}}{N_{\rm dir}}
 \simeq 1.18 .
 \label{hyper_ratio0}
\end{equation}
The sequential contribution is consequently not a small
correction to the direct channel; the two pathways are of
comparable magnitude in this calculation.  Hypertriton
production has also been measured in Pb--Pb collisions by the
ALICE Collaboration \cite{ALICEHypertriton2016}.

The same calculation provides an instructive indication of the
different temporal sensitivities of the two formation channels.
After a free-streaming interval of about
$\Delta t=35$ fm/$c$, the direct and sequential contributions
are approximately
\begin{equation}
 N_{\rm dir}(35~{\rm fm}/c)
 \simeq 4.4\times10^{-5},
 \qquad
 N_{\rm seq}(35~{\rm fm}/c)
 \simeq 7.3\times10^{-5},
 \label{hyper_yields35}
\end{equation}
so that the total abundance changes relatively moderately,
whereas the composition of the formation channels changes more
substantially.  In particular,
\begin{equation}
 R_{\rm seq/dir}(35~{\rm fm}/c)
 \simeq
 \frac{7.3}{4.4}
 \simeq 1.66 ,
 \label{hyper_ratio35}
\end{equation}
compared with $R_{\rm seq/dir}(0)\simeq1.18$.

For orientation, the time dependence of these two contributions
may be summarized phenomenologically by
\begin{equation}
 N_i(\Delta t)
 =
 N_i(0)
 \exp\left(-\frac{\Delta t}{\tau_i^{\rm eff}}\right).
 \label{hyper_effective}
\end{equation}
Using the above benchmark values gives approximately
\begin{equation}
 \tau_{\rm dir}^{\rm eff}
 \simeq 8.3\times10^{1}\ {\rm fm}/c,
 \qquad
 \tau_{\rm seq}^{\rm eff}
 \simeq 4.4\times10^{2}\ {\rm fm}/c .
 \label{hyper_times}
\end{equation}
These quantities should not be identified directly with the
relaxation eigenvalues of the memory kernel in Eq.~(\ref{eq:multiscale}).
They are effective survival scales extracted from the time
dependence of the coalescence calculation and are quoted here
only to characterize the markedly different temporal
sensitivities of the two formation pathways.

The more robust quantity for the present purpose is therefore
the ratio $R_{\rm seq/dir}$.  The above benchmark corresponds
to
\begin{equation}
 \frac{
 R_{\rm seq/dir}(35~{\rm fm}/c)}
 {R_{\rm seq/dir}(0)}
 \simeq 1.4 .
 \label{hyper_ratio_change}
\end{equation}
Thus a change of only about $20\%$ in the total hypertriton
abundance can coexist with an approximately $40\%$ change in
the relative composition of the two formation channels.  This
provides a quantitative illustration of the general statement
that a final abundance need not uniquely determine the
dynamical path by which it was produced.

The reduced-dynamical interpretation leads to a further
prediction.  Since the sequential channel contains an
additional dynamical stage whose response is encoded in
$G_d(t,t')$, its relative contribution should depend on the
duration and expansion history of the hadronic stage.  Defining
\begin{equation}
 R_{\rm seq/dir}
 =
 \frac{N_{d+\Lambda}}
      {N_{p+n+\Lambda}},
 \label{hyper_Rprediction}
\end{equation}
the minimal memory picture predicts that this ratio should not
in general be independent of the effective evolution time.
In a regime in which the sequential contribution retains its
correlations more efficiently than the direct contribution, one
expects schematically
\begin{equation}
 \tau_{\rm macro}\ \uparrow
 \qquad\Longrightarrow\qquad
 R_{\rm seq/dir}\ \uparrow .
 \label{hyper_prediction}
\end{equation}
The benchmark values above suggest that an order-$40\%$ change
in this ratio over an evolution interval of order
$35$ fm/$c$ is dynamically plausible.  The precise centrality
dependence, however, requires a realistic treatment of the
expansion, breakup, regeneration, and medium-dependent
formation rates and is not predicted quantitatively by the
present reduced model.

More generally, if the unresolved formation sector contains
several independent relaxation modes, Eq.~(\ref{eq:multikernel}) gives
\begin{equation}
 K_H(\Delta t)
 \simeq
 \sum_i A_i
 \exp\left(-\frac{\Delta t}{\tau_i}\right).
 \label{hyper_multiscale}
\end{equation}
The appearance of several exponentials is therefore associated
with several relaxation eigenmodes of the eliminated dynamics,
rather than merely with the existence of several reaction
pathways.  The direct--sequential hypertriton system provides a
particularly useful testing ground because the sequential
pathway makes the role of an additional intermediate dynamical
sector explicit.

The same reasoning is not restricted to exponential relaxation.
The generic reduced equation, Eq.~(\ref{eq:SK_memory}), allows qualitatively
different temporal structures of the effective kernel.
Schematically, one may distinguish
\begin{equation}
\begin{array}{lll}
 K(\Delta t)\propto e^{-\Delta t/\tau},
 &\qquad& {\rm single\mbox{-}relaxation},\\[2mm]
 K(\Delta t)\propto
 \displaystyle\sum_i A_i e^{-\Delta t/\tau_i},
 &&
 {\rm multi\mbox{-}relaxation},\\[2mm]
 K(\Delta t)\propto(\Delta t)^{-\alpha},
 &&
 {\rm long\mbox{-}memory},\\[2mm]
 K(\Delta t)\propto
 e^{-\gamma\Delta t}\cos(\omega\Delta t),
 &&
 {\rm oscillatory\mbox{-}memory}.
\end{array}
\label{kernel_classes}
\end{equation}
The microscopic origins of these kernels may be entirely
different, but their temporal structures provide a natural
classification at the level of reduced nonequilibrium dynamics.

This leads to a testable extension of the universality proposed
in this work.  Different physical systems may be classified not
only by their microscopic degrees of freedom, but also by the
temporal spectrum of the memory kernel generated under
dynamical reduction.  Conversely, measurements sensitive to
formation chronology and correlations may in principle be used
to infer properties of dynamical sectors that are no longer
explicitly present in the final state.  In this sense,
comparison among different composite species could provide a
form of ``memory-kernel spectroscopy'' of their nonequilibrium
formation dynamics.

The heavy-ion and FMT examples then represent two distinct
members of this broader classification.  The minimal heavy-ion
system produces a finite-lifetime exponential kernel, whereas
the FMT system produces an expansion-dressed gravitational
kernel through the
$\rho_\phi\to\rho_l\to H\to\rho_\phi$ feedback loop.  The
prediction is that other systems containing eliminated
dynamical sectors with non-negligible response times should
generate kernels whose temporal form reflects their
characteristic relaxation spectrum.  Identifying such
structures in additional physical systems would provide a
direct test of whether the universality represented by
Eq.~(\ref{eq:SK_memory}) extends beyond the two realizations studied here.
\section{Discussion and conclusion}

We have examined light-nuclei formation in heavy-ion
collisions and cosmological Bose--Einstein condensation
from the common viewpoint of reduced nonequilibrium
dynamics.  The central result is that these microscopically
very different systems provide two concrete realizations
of the same causal reduced structure represented by
Eq.~(\ref{eq:SK_memory}).  The universality discussed here therefore
resides neither in the identity of the microscopic degrees
of freedom nor in an identical cascade topology, but in
the dynamical structure that remains after an additional
coupled sector has been eliminated.

In the heavy-ion realization, the additional sector is
represented in the minimal model by the finite-lived
$\Delta$ component.  Eliminating this component gives
Eqs.~(\ref{eq:HI_memory}) and (\ref{eq:HI_kernel}), and its finite lifetime
$\tau_\Delta=\Gamma_\Delta^{-1}$ appears explicitly as
the characteristic memory time of the reduced nucleon
dynamics.  Figures~1 and 2 illustrate the corresponding
time ordering and its sensitivity to this additional
dynamical time scale.  The role of the minimal model is
not to establish a unique $\Delta$-mediated mechanism for
the ALICE data, but to exhibit explicitly how a finite-lived
hadronic sector generates a causal memory term under
dynamical reduction.

The FMT realization is microscopically different.  The
localized component $\rho_l$ is generated from the
coherent component $\rho_\phi$ and is therefore not a
precursor analogous to the $\Delta$ resonance.  However,
because $\rho_l$ contributes to the Hubble rate and the
Hubble rate in turn enters the evolution of $\rho_\phi$,
the system contains the feedback loop
\[
 \rho_\phi\longrightarrow\rho_l
 \longrightarrow H\longrightarrow\rho_\phi .
\]
Eliminating the localized sector therefore feeds its
previous evolution back into the coherent dynamics.
Eqs.~(\ref{delta_rhol}) and (\ref{eq:FMT_memory}) demonstrate explicitly that this
reduction generates a causal memory kernel.  In this case
the memory is mediated by gravitational expansion rather
than by the direct decay of a finite-lived resonance.

The two memory kernels are neither identical nor expected
to be so.  Their different forms encode the microscopic
physics of the two systems.  What survives these
differences is the operation of dynamical reduction:
when a coupled sector has a response time that is not
negligible on the time scale of the retained dynamics,
eliminating that sector leaves a causal record of its
previous evolution.  When the response time is much
shorter than the macroscopic evolution time, this memory
becomes effectively local and the usual Markovian
description is recovered.

This gives a precise meaning to the universality proposed
in the present work.  The heavy-ion and cosmological
systems need not share the same particles, interactions,
absolute time scales, expansion laws, or microscopic
sequence of processes.  They belong to the same
reduced-dynamical class because coarse graining over a
dynamically relevant sector leads in both cases to the
causal structure of Eq.~(\ref{eq:SK_memory}).  The heavy-ion system
realizes this structure directly through the finite
lifetime of an intermediate hadronic component, whereas
the FMT system realizes it indirectly through the
$\rho_\phi\to\rho_l\to H\to\rho_\phi$ feedback loop.
This establishes the concrete connection between the two
physical examples considered here and the general
Schwinger--Keldysh reduced-dynamics framework of
Ref.~[9].

An important consequence is that an equilibrium-like
final state does not uniquely determine the dynamical
path by which it was reached.  Coarse graining can hide
intermediate or auxiliary dynamical sectors even when
final observables admit a successful statistical or
equilibrium parametrization.  Conversely, the presence
of such nonequilibrium dynamics does not preclude
equilibrium-like regularities in final observables.
Temporal ordering, correlations, response times, and
memory effects therefore contain information that is not
encoded in the final-state abundance alone.
The hypertriton application provides a concrete realization
of this consequence and leads to an experimentally testable
prediction.  In the minimal description, the direct
$p+n+\Lambda$ channel acts as a local source, whereas the
sequential $p+n\to d$ followed by $d+\Lambda$ channel
contains an additional intermediate dynamical sector.
Eliminating the deuteron sector therefore generates a
retarded contribution to the reduced hypertriton dynamics.
The benchmark calculation of Ref.~\cite{ZhangKo2018}
illustrates the possible magnitude of this effect: while
the total hypertriton abundance changes by only about
$20\%$ over an evolution interval of about $35~{\rm fm}/c$,
the ratio of sequential to direct contributions changes by
about $40\%$, as shown in Eq.~(\ref{hyper_ratio_change}).
This numerical value should be regarded as a benchmark
rather than as an absolute prediction of the present
reduced model.  The prediction of the memory picture is
instead the systematic dependence
\[
 \tau_{\rm macro}\uparrow
 \quad\Longrightarrow\quad
 R_{\rm seq/dir}\uparrow
\]
in the regime where the sequential contribution retains
its correlations more efficiently than the direct one.
Systematic measurements under collision conditions with
different effective hadronic evolution histories can
therefore test whether the relative formation-channel
composition retains dynamical information that is hidden
in the total final yield.
For light-nuclei production, a quantitative comparison
with statistical hadronization, coalescence, and
microscopic transport requires a more complete treatment
of inverse reactions, medium-dependent rates,
regeneration, and hydrodynamic evolution.  Likewise, the
quantitative magnitude of the FMT memory depends on the
cosmological background and on the strength of the
gravitational feedback.  These features determine the
strength of each particular realization, but not the
common reduced-dynamical principle identified here.

The common lesson of the systems studied here is that
dynamical information can survive coarse graining even
when it is no longer visible in the final abundance.
In both the heavy-ion and cosmological examples, an
additional dynamical sector stores information about the
previous evolution and feeds it back into the retained
degrees of freedom.  After that sector is eliminated,
this information appears as a causal memory term.
The microscopic origin of the memory is different in the
two systems---a finite-lived hadronic sector in one case
and gravitational feedback in the other---but the reduced
dynamical structure is the same.  This is the universality
identified in the present work.  The hypertriton example
further illustrates its observable consequence: the total
final yield may change only moderately while the relative
weights of different formation pathways retain a much
stronger dependence on the preceding evolution.  Thus,
observables sensitive to formation history, rather than
final abundances alone, provide a natural way to test the
dynamical memory discussed here.
\begin{acknowledgments}
This work was supported by JSPS KAKENHI Grant Number JP25H00653.
\end{acknowledgments}


\end{document}